\documentclass{elsart}
\usepackage{epsfig}
\usepackage{amssymb}

\begin{document}
\begin{frontmatter}

\title{The Peter Principle Revisited: A Computational Study} 

\author
{Alessandro Pluchino} \ead{alessandro.pluchino@ct.infn.it}
\address{Dipartimento di Fisica e Astronomia,  Universit\'a di Catania,\\
and INFN sezione di Catania,  Via S. Sofia 64, I-95123 Catania, Italy} 

\author
{Andrea Rapisarda} \ead{andrea.rapisarda@ct.infn.it}
\address{Dipartimento di Fisica e Astronomia,  Universit\'a di Catania,\\
and INFN sezione di Catania,  Via S. Sofia 64, I-95123 Catania, Italy} 

\author
{Cesare Garofalo}  \ead{cesaregarofalo@yahoo.com}
\address {Dipartimento di Sociologia e Metodi delle Scienze Sociali, Universit\'a di Catania, \\
Via Vittorio Emanuele II 8, I-95131 Catania, Italy}

%
%

\begin{abstract}


In the late sixties the Canadian psychologist Laurence J. Peter advanced an apparently paradoxical principle, named since then after him, which can be summarized as follows: {\it 'Every new member in a hierarchical organization climbs the hierarchy until he/she reaches his/her level of maximum incompetence'}. Despite its apparent unreasonableness, such a principle would realistically act in any organization where the mechanism of promotion rewards the best members and where the mechanism at their new level in the hierarchical structure does not depend on the competence they had at the previous level, usually because the tasks of the levels are very different to each other. Here we show, by means of agent based simulations, that if the latter two features actually hold in a given model of an organization with a hierarchical structure, then not only is the Peter principle unavoidable, but also it yields in turn a significant reduction of the global efficiency of the organization. Within a game theory-like approach, we explore different promotion strategies and we find, counterintuitively, that in order to avoid such an effect the best ways for improving the efficiency of a given organization are either to promote each time an agent at random or to promote randomly the best and the worst members in terms of competence.

\end{abstract}

\begin{keyword}
Peter Principle, Organizations Efficiency, Agent Based Models
\end{keyword}
\end{frontmatter}

\section{Introduction: The Peter Principle}

The efficiency of an organization in terms of improving the ability to perform a job minimizing the respective costs is a key concept in several fields like economics \cite{Farmer} and game theory \cite{Gibbons}. But it could also be very important in ecology to understand the behaviour of social insects \cite{Dornhaus}, in computer science when you have to allocate different tasks to a cluster of computers having different performances \cite{Foster} or in science policy concerning how individual tasks are distributed among the thousands of members of a big collaboration, like those working for example at a large collider. 
Common sense has always been widely used in any hierarchical organization to manage the system of promotions: it tells us that a member who is competent at a given level, will be competent also at an higher level of the hierarchy, so it seems a good deal, as well as a meritorious action, to promote such a member to the next level in order to ensure the global efficiency of the system. 
The problem is that common sense, in many areas of our everyday life, often deceives us. In 1969 the Canadian psychologist Laurence J. Peter warned that the latter statement could be true also for the promotions management in a hierarchical organization \cite{Peter}. 
\\
Actually, the simple observation that a new position in the organization requires different work skills for effectively performing the new task (often completely different from the previous one), could suggest that the competence of a member at the new level could not be correlated to that at the old one. Peter speculated that we may consider this new degree of competence as a random variable, even taking into account any updating course the organization could require before the promotion: this is what we call the Peter hypothesis. If the Peter hypothesis holds, and if one promotes each time the most competent member at the involved level, it could turn out a paradoxical process for which competent members will climb up the hierarchical ladder indefinitely, until they reach a position where they will be no longer competent and therefore no longer promoted. This is the so called Peter principle, whose long term consequence seems to imply an unavoidable spreading of the incompetence over all of the organization and would be in danger of causing a collapse in its efficiency, as also confirmed already in 1970 by a mathematical analysis of J.Kane \cite{Kane}.
\\
More recently several reflections on bureaucratic inefficiency have been carried out in the context of social science, politics and business management  \cite{Adams,Lazear,Dickinson,Klimek1,Klimek2,Buchanan1}, some of which were directly inspired by the Peter principle and with the purpose of circumvent its adverse effects. 
However, as far as we know, we still lack a computational study that not only would reproduce the Peter principle dynamics, but also would allow, in particular, the exploration of alternative strategies in order to find the best way for improving the efficiency of a given organization \cite{Note}.
\\
In the last few years the help of hard sciences, like physics and mathematics, has been frequently advocated in order to get a more quantitative understanding of social sciences mechanisms \cite{Nature,Buchanan2,Hedstrom1,Hedstrom2}. It is now largely accepted that simple models and simulations inspired by statistical physics are able to take into account collective behaviour of large groups of individuals, discovering emergent features independent of their individual psychological attributes and very often counterintuitive and difficult to predict following common sense \cite{Epstein1,Miller,Castellano}. Along these lines, by means of an agent based simulation approach \cite{Epstein2,Gilbert,Wilensky}, here we study the Peter principle process within a general context where different promotions strategies compete one with one another for maximizing the global efficiency of a given hierarchical system. 

\begin{figure}  
\begin{center}
\epsfig{figure=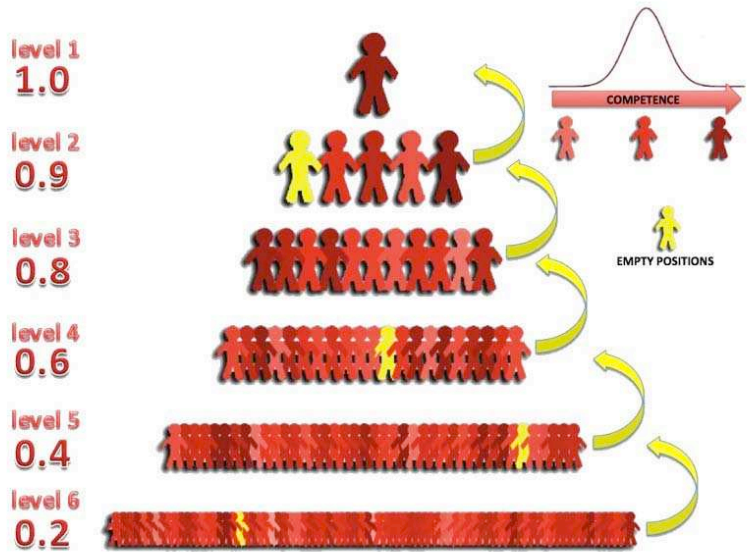,width=11truecm,angle=0}
\end{center}
\caption{ {\it Schematic view of a hierarchical pyramidal organization}. We consider here an organization with 160 positions divided into six levels. Each level has a different  number of members (which decreases climbing the hierarchy) with a different responsibility, i.e. with a different weight on the global efficiency of the organization, reported on the left side. The member colour indicates the degree of competence, which at the beginning is normally distributed with average 7.0 and variance 2.0. Empty positions are in yellow (see the text). 
}
\label{organization}
\end{figure}

\section{Dynamical Rules of the Model}

In order to simplify the problem, we chose for our study a prototypical pyramidal organization, (see Fig.\ref{organization}), made up of a total of 160 positions distributed over six levels numbered from 6 (the bottom level) to 1 (the top one), with 81 members (agents) in level 6, 41 in level 5, 21 in level 4, 11 in level 3, 5 in level 2 and 1 in level 1. We verified that the numerical results that we found for such an organization are very robust and show only a little dependence on the number of levels or on the number of agents per level (as long as it decreases going from the bottom to the top).  
Each agent is characterized only by an {\it age} and by a {\it degree of competence}. The degree of competence, which includes all the features (efficiency, productivity, care, diligence, ability to acquire new skills) characterizing the average performance of an agent in a given position at a given level, etc., is a real variable with values ranging from 1 to 10 and is graphically represented with a colour scale with increasing intensity. The age, however, is an integer variable included in the range 18-60, which increases by one unit per each time step. 
\\
The snapshot reported in Fig.\ref{organization} shows, as an example, a given realization of the initial conditions, where both the competence and the age of each agent have been selected randomly inside two appropriate normal distributions with, respectively, means 7.0 and 25 and standard deviations 2.0 and 5.
At each time step all the agents with a competence under a fixed dismissal-threshold or with an age over a fixed retirement-threshold leave the organization and their positions become empty, while their colour becomes yellow (the dismissal-threshold is arbitrarily fixed to 4 and the retirement-threshold to an age of 60). Simultaneously, any empty position at a given level is filled by promoting one member from the level immediately below, going down progressively from the top of the hierarchy until the bottom level has been reached. Finally, empty positions at the bottom level are filled with the recruitment of new members with the same normal distribution of competences described before.
\\
We consider two possible mechanisms of transmission of competence of an agent from one level to the next one: the {\it common sense hypothesis}, where a member inherits his/her old competence in his/her new position with a small random variation $\delta$ (where $\delta$ can assume random values included within $\pm10\%$ of the maximum value in the competence scale, i.e. $\delta \in [-1,1]$), and the {\it Peter hypothesis}, where the new competence of every agent is independent of the old one and is assigned randomly (again with the same normal distribution than before). For each one of these two cases we take into account three different ways of choosing the agent to promote at the next level: the most competent ({\it The Best} strategy, suggested by common sense and adopted also in the Peter principle), the least competent ({\it The Worst} strategy) or one agent at random (the {\it Random} strategy). 
\\
At this point, in order to evaluate the global performance of the organization, we introduce a parameter, called {\it global efficiency}, which is calculated by summing the competences of the members level by level, multiplied by a level-dependent factor of responsibility ($r_i$, with $i=1,2,$...$,6$) ranging from 0 to 1 and increasing on climbing the hierarchy (such a factor, shown in the left side of Fig.\ref{organization}, takes into account the weights that the performances of the agents of different levels have in the global efficiency of the organization). Finally, the result is normalized to its maximum possible value ($Max(E)$) and to the total number of agents ($N$), so that the global efficiency ($E$) can be expressed as a percentage. Therefore, if $C_i$ is the total competence of level $i$, the resulting expression for the efficiency is $E(\%)=\frac{\sum_{i=1}^6 C_i r_i}{Max(E) \cdot N} \cdot 100$, where $Max(E)=\sum_{i=1}^6 10 \cdot n_i r_i / N$ (being $n_i$ the number of agents of level $i$).

\begin{figure}  
\begin{center}
\epsfig{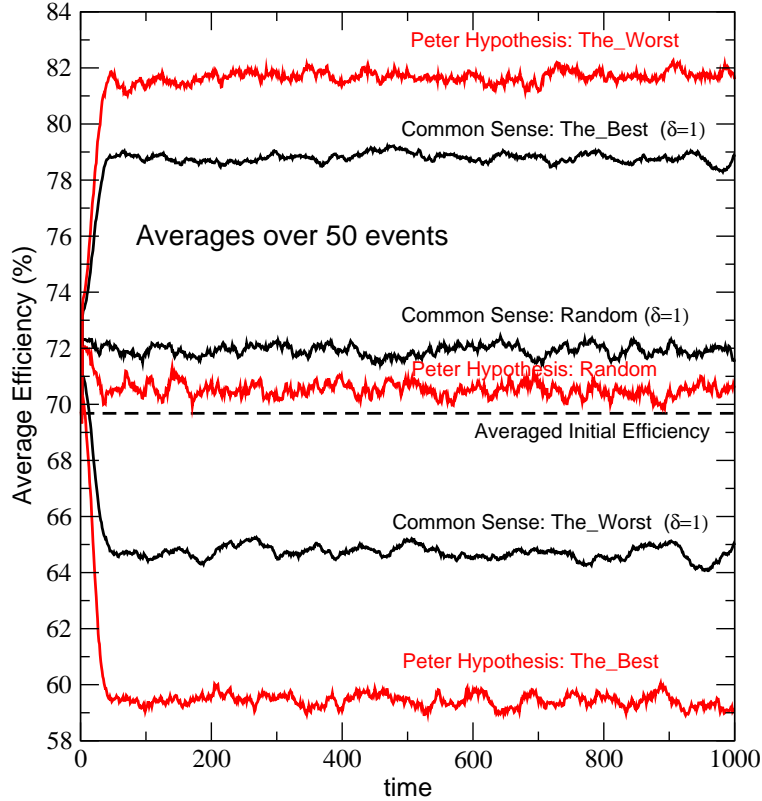}
\end{center}
\caption{ {\it Efficiency time evolution for different strategies}. This figure shows the time evolution of the global efficiency averaged over 50 realizations of the initial competence distribution (events). Starting from the same initial average value (dashed line), the efficiency is plotted for 1000 time steps for the six combinations of the CS and PH mechanisms with the three different promotion strategies ({\it The Best}, {\it The Worst} and {\it Random}). 
}
\label{global-efficiency}
\end{figure}

\section{Strategies in Competition: Simulations Results}

We realized all the simulations presented in the paper with NetLogo \cite{Wilensky}, a programmable environment designed for developing agent based simulations of complex systems.  
In Fig.\ref{global-efficiency} we show the time evolution of the global efficiency considering the six possible combinations among the mechanisms of competence transmission and the promotion strategies. 
The evolution is calculated for 1000 time steps, a duration long enough to reach a stationary (on average) asymptotic value, and is further averaged over 50 different realizations of the initial conditions. The corresponding standard deviations are $\sim1\%$  and are not reported. The simulations start always from the same 50 initial configurations of competences, so the initial average efficiency is fixed at the value $69.68\%$ (dashed line). At  t=0 all the curves seem to start from a point slightly above this line because the initial random distribution of competences produces many empty positions which are immediately filled in the first few steps then producing, regardless of the other parameters, a sudden small initial increment of about $2\%$ in the global average efficiency. We verified that all the results presented do not depend drastically on small changes in the value of the free parameters: therefore the emerging scenario and the corresponding conclusions appear very robust.
\begin{figure}  
\begin{center}
\epsfig{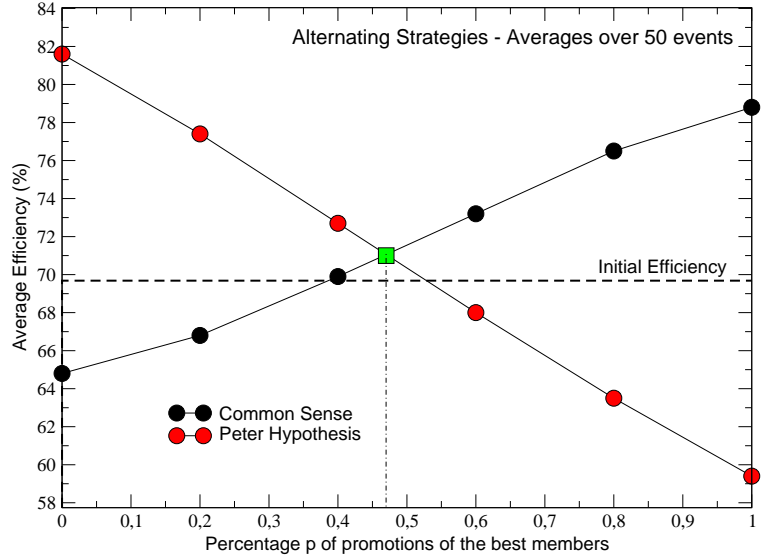}
\end{center}
\caption{ {\it Efficiency for alternating strategies}. The plot shows the asymptotic values of global efficiency averaged over the last 400 time steps and over 50 events in the case of an alternating {\it Best-Worst} strategy of promotions, as a function of the percentage $p$ of promotions of the best members with respect to all the promotions and for both the CS and PH mechanisms of competence transmission. In correspondence with the values p=0.0 and p=1.0 (respectively {\it The Worst} and {\it The Best} strategies) one recovers the asymptotic values of the previous figure for both scenarios. The point [0.47,71] (green square) gives the most convenient strategy to adopt if one does not know which mechanism of competence transmission is operative in the organization. 
}
\label{alternating-strategies}
\end{figure}
\\
Looking at Fig.\ref{global-efficiency} it is clearly apparent that if one always promotes the best member, as is usually done by all real organizations, the asymptotic value of the average efficiency (AE) significantly increases ($+9\%$) with respect to the initial efficiency only if common sense (CS) transmission occurs: if, on the contrary, one assumes as valid the Peter hypothesis (PH), a significant decrement of AE occurs ($-10\%$) as intuitively predicted by Peter. Since in general it is difficult to know which mechanism is actually operative in a given real organization, that of promoting the best member does not turn out to always be a winning strategy! Let us consider the opposite strategy, i.e. promoting the worst member. Again the resulting AE strictly depends on the transmission mechanism, but in the opposite way: in fact, in this case the strategy is a winning one for PH ($+12\%$), while it is a losing one for CS ($-5\%$). Finally, the third strategy Ð that of promoting one agent at random Ð gives more similar results in both cases, although the improvement of the initial efficiency is limited ($+2\%$ for CS, $+1\%$ for PH). 
\\
In order to obtain exactly the same efficiency for CS and PH, in Fig.\ref{alternating-strategies} we introduce a fourth {\it Best-Worst} strategy, where the best and the worst agents are chosen alternately with a variable percentage $p$. For $p=0$ and $p=1$ we recover respectively {\it The Worst} and {\it The Best} cases shown in Fig.\ref{global-efficiency}, while for values $p \in ]0,1[$ all the intermediate situations between these limiting two cases are obtained. Interpolating the numerical results for the CS and PH scenarios, we found that an alternating strategy with an almost random choice between the best and the worst members ($p=0.47$; see the green square) produces on average the same asymptotic value of global efficiency in the two cases, with a limited gain ($+1.5\%$) but without losses. 
\\
We summarize in Table 1 the percentages of gain or loss obtained for the different strategies applied. These results confirm that, within a game theory-like approach, if one does not know what mechanism of competence transmission is operative in a given organization, the best promotion strategy seems to be that of choosing a member at random or, at least, that of choosing alternately, in a random sequence, the best or the worst members. This result is quite unexpected and counterintuitive, since the common sense tendency would be that of always promoting the best member, a choice that, if the Peter hypothesis holds, turns out to be completely wrong. On the other hand, by applying one of the two strategies {\it Random} and {\it Best-Worst}, losses can be successfully avoided without any further (possibly expensive) precaution of the organization's managers (such as specialization or updating courses).  
\begin{table} 
\begin{center}
\begin{tabular}{c c c c c}
\hline\hline
    & {\it The Best} & {\it The Worst} & {\it Random} & {\it Best-Worst (p=0.47)} \\
\hline
{\bf Common Sense} & $+9\%$ & $-5\%$ & $+2\%$ & $+1.5\%$ \\
{\bf Peter Hypothesys} & $-10\%$ & $12\%$ & $+1\%$ & $+1.5\%$\\
\hline\hline
\end{tabular}
\caption{{\it Comparison of gains and losses for all the strategies adopted}.The table summarizes the average gain and loss percentage calculated with respect to the average initial efficiency by following the different promotion strategies. The table indicates that {\it The Best} and {\it The Worst} strategies provide at the same time significant gains and consistent losses, depending on the mechanisms of competence transmission at stake (common sense or the Peter hypothesis); therefore the best strategies to adopt when such a mechanism is unknown are clearly the {\it Random} and the alternating {\it Best-Worst} ones (the latter, for $p=0.47$, gives also exactly the same positive gain for the two scenarios). }
\end{center}
\label{summary-table}
\end{table}
\\
Finally, in Fig.\ref{summary-table} we quantitatively verified the apparently surprising statement of the Peter principle, i.e. the fact that each member in a hierarchical organization climbs the hierarchy until he/she reaches the level where his/her competence is minimal. We found that no matter how many levels an agent crosses in his/her career: if one adopts the strategy of promoting the best member and if the PH holds, then all the members will end their career at the level where their competence is minimal or, what is the same, where their incompetence is the greatest. Therefore PeterÕs intuition is definitively correct: this dangerous mixture yields the rapid decrease of efficiency observed in Fig.\ref{global-efficiency}. It is interesting to notice that if instead the PH is combined with the strategy of promoting the worst member, then the situation is diametrically opposite, i.e. each agent climbs the hierarchy until he/she reaches the level where his/her competence is at the maximum. Finally, the other PH lines in Fig.\ref{summary-table}, corresponding to the random promotion strategy, show constant competences during all the careers, and the same happens in the analogous case with the CS mechanism.
 
\begin{figure}  
\begin{center}
\epsfig{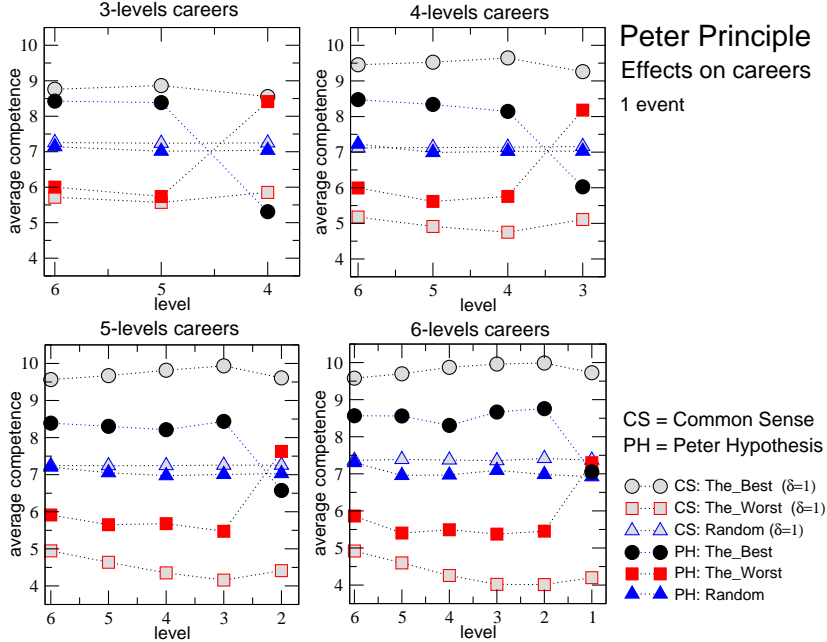}
\end{center}
\caption{ {\it Career evolutions}. We show in the figure the behaviour of the competence per level calculated while each agent climbs the hierarchy for the six different cases of Fig. \ref{global-efficiency}. It is the computational verification of the Peter statement: 'Everynew member in a hierarchical organization climbs the hierarchy until he/she reaches his/her level of maximum incompetence'. Each point is an average over all the agents who, in a single event, go up from the sixth, bottom level to one of the upper levels. As expected, the Peter effect clearly appears only in the PH cases with the strategy of promoting the best agent (full black circles). Furthermore it does not depend on the number of levels that the agents cross in their careers.
}
\label{careers-evolution}
\end{figure}

\section{Conclusions}

In conclusion, our computational study of the Peter principle process applied to a prototypical organization with pyramidal hierarchical structure shows that the strategy of promoting the best members in the PH case induces a rapid decrease of efficiency, while it works well only if members would ideally maintain their competence at each level, an hypothesis that, although in agreement with common sense, seems in practice very unrealistic in the majority of the real situations. On the other hand we obtained the counterintuitive result that the best strategies for improving, or at least for not diminishing, the efficiency of an organization, when one ignores the actual mechanism of competence transmission, are those of promoting an agent at random or of randomly alternating the promotion of the best and the worst members. We think that these results could be useful to guide the management of large real hierarchical systems of different natures and in different fields \cite{Applet}.


\begin{thebibliography}{00}

\bibitem{Farmer} J.D.Farmer, A.W.Lo, {\it Frontiers of finance: Evolution and efficient markets}, Proc. Natl. Acad. Sci. USA {\bf 96}, 9991-9992 (1999) 

\bibitem{Gibbons}  R.Gibbons, {\it Game Theory for Applied Economists}, Princeton University Press, 1992

\bibitem{Dornhaus} 	A.Dornhaus, {\it Specialization does not predict individual efficiency in an ant}, PLoS Biology {\bf 6}, 2368 (2008)

\bibitem{Foster} I.Foster, C.Kesselman, S.Tuecke, {\it The Anatomy of the Grid: Enabling Scalable Virtual Organizations}, Intl. J. Supercomputer Applications (2001) 

\bibitem{Peter} L.J.Peter, R.Hull, {\it The Peter Principle: why things always go wrong}, New York - William Morrow and Company (1969)

\bibitem{Kane} J.Kane, {\it Dynamics of the Peter Principle}, Management Science {\bf 16}, 1970, B-800-B-811

\bibitem{Adams} S.Adams, {\it The Dilbert Principle}, HarperBusiness (1996)

\bibitem{Lazear} E.P.Lazear, {\it The Peter Principle: A Theory of Decline}, Journal of Political Economy {\bf 112(1)}, S141-S163 (2001)

\bibitem{Dickinson} D.L.Dickinson and M.Villeval, {\it The Peter Principle: An Experiment}, GATE Working Paper, No. 07-28 (2007)

\bibitem{Klimek1} P.Klimek et al., {\it Parkinson's Law quantified: three investigations on bureaucratic inefficiency}, J. Stat. Mech. P03008 (2009)

\bibitem{Klimek2} P.Klimek et al., {\it To how many politicians should government be left?}, Physica A, {\bf 388} (2009) 3939-3947

\bibitem{Buchanan1} M.Buchanan, {\it Explaining the curse of work}, New Scientist, issue 2690 (2009)

\bibitem{Note}  The fact that the Peter Principle is underestimated in Social Sciences is confirmed also, for example, by the absence of entries for  'Peter Principle' in {\it 'The New Palgrave Dictionary of Economics'} and in the {\it 'International Encyclopedia of the Social $\&$ Behavioral Sciences'}, which are two very important texts in this field. 

\bibitem{Nature} See for example the editorial Nature  435, 1003 (2005).

\bibitem{Buchanan2} M.Buchanan, Nature Physics 4, 159 (2008).

\bibitem{Hedstrom1} P.Hedstrom, R.Swedberg (Ed.), {\it Social Mechanisms: An Analytical Approach to Social Theory}, Cambridge University Press (1998)

\bibitem{Hedstrom2} P. Hedstrom, {\it Dissecting the Social: On the Principles of Analytical Sociology}, Cambridge University Press (2005)

\bibitem{Epstein1} J.Epstein, R.Axtell, {\it Growing Artificial Societies: Social Science from the Bottom Up}, MIT Press (1996)

\bibitem{Miller} 	J.H.Miller, S.Page, {	\it Complex Adaptive Systems: An Introduction to Computational Models of Social Life}, Princeton University Press (2007)

\bibitem{Castellano} C.Castellano, S.Fortunato, V.Loreto, {\it Statistical physics of social dynamics}, Rev. of Mod.Phys., Vol. {\bf 81} (2009)

\bibitem{Epstein2} J.M.Epstein (Ed.), {\it Generative Social Science: Studies in Agent-Based Computational Modeling}, Princeton University Press (2007)

\bibitem{Gilbert}	N.Gilbert, {\it Agent-Based Models (Quantitative Applications in the Social Sciences)}, Sage Publications (2007)

\bibitem{Wilensky} U.Wilensky Ð NetLogo Ð http://ccl.northwestern.edu/netlogo. Center for Connected Learning and Computer-Based Modeling. Northwestern University, Evanston, IL (1999)

\bibitem{Applet}
For users that would like to verify our findings, a Java applet of the NetLogo program used to perform the simulations presented in this paper is available at the following web address: 

$http://www.ct.infn.it/cactus/peter\_principle\_sup\_material.html $.

\end{thebibliography}
\end{document}